\documentclass[conference, a4paper, 10pt]{IEEEtran}
\usepackage{amsfonts}
\IEEEoverridecommandlockouts

\ifCLASSINFOpdf
\else
\fi
\usepackage{epsfig}
\usepackage{graphicx}
\usepackage{psfig}
\usepackage{subfigure}
\usepackage{epsf}
\usepackage[cmex10]{amsmath}
\usepackage{booktabs}
\usepackage{fancyhdr}

\usepackage{color}
\usepackage[letterpaper, left=0.625in, right=0.625in, bottom=1in, top=0.70in]{geometry}
\newcommand{\bc}{\begin{center}}
	\newcommand{\ec}{\end{center}}
\newcommand{\be}{\begin{equation}}
\newcommand{\ee}{\end{equation}}
\newcommand{\bea}{\begin{eqnarray}}
\newcommand{\eea}{\end{eqnarray}}

\renewcommand\IEEEkeywordsname{Index Terms}

\begin{document}
	
	\title{Energy-Efficient Design for IRS-Assisted MEC Networks with NOMA}
	
		\author{Qun~Wang\textsuperscript{\dag}, Fuhui~Zhou\textsuperscript{\ddag}, Han Hu$^*$,  Rose Qingyang Hu\textsuperscript{\dag}\\ 
		\textsuperscript{\dag}Department of Electrical and Computer Engineering, Utah State University, Logan, UT, USA\\%
		\textsuperscript{\ddag}College of Electronic and Information Engineering,\\
		 Nanjing University of Aeronautics and Astronautics, Nanjing, China\\
		$^*$Jiangsu Key Laboratory of Wireless Communications, \\ Nanjing University of Posts and Telecommunications, Nanjing, China\\
		Emails: claudqunwang@ieee.org, zhoufuhui@ieee.org, han\_h@njupt.edu.cn, rose.hu@usu.edu
	}

	\maketitle
	
	\IEEEpeerreviewmaketitle
	\begin{abstract}
Energy-efficient design is of crucial importance in wireless internet of things (IoT) networks. In order to serve massive users while achieving an energy-efficient operation, an intelligent reflecting surface (IRS) assisted mobile edge computing (MEC) network with non-orthogonal multiple access (NOMA) is studied in this paper. The energy efficiency (EE) is maximized by jointly optimizing the offloading power, local computing frequency, receiving beamforming, and IRS phase-shift matrix. The problem is challenging to solve due to the non-convex fractional objective functions and the coupling among the variables. A semidefinite programming relaxation (SDR) based alternating algorithm is developed.
Simulation results demonstrate that the proposed design outperforms the benchmark schemes in terms of EE. Applying IRS and NOMA can effectively improve the performance of the MEC network.
\end{abstract}
\IEEEpeerreviewmaketitle
\section{Introduction}
The launch of fifth-generation (5G) wireless communication networks leads to an explosive increment of internet of things (IoT) devices, which will generate massive data  that need to be processed timely either through local or remote computing facilities. Limited Bandwidth and limited  device computation capability  have gradually become a bottleneck to meet delay requirement for the massive amount of data. 
Meanwhile, the research on sixth-generation (6G) wireless networks has taken off and has been attracting ever increasing attention from academia and industry \cite{6g}.  Future wireless networks aim for realizing ultra-high spectrum efficiency (SE) and energy efficiency (EE), ultra-dense user connectivity, and ultra  low latency. EE is  critically important in future wireless   networks since massive energy constrained IoT devices need to be reliably connected to support smart city, smart factory, etc. Moreover, it is imperative to address the finite computation capacity issue of IoT devices.


Mobile edge computing (MEC) is envisioned as a promising paradigm by exploiting the edge servers that are deployed close to the end-users. In wireless networks, edge nodes, such as base stations and edge routers, can be equipped with high computing and storage capabilities.  MEC enables users to offload their tasks to edge servers for processing. It is important to have efficient offloading in MEC as massive local devices may need to offload tasks to edge servers for computing. Towards that end, NOMA can be applied to further enhance EE and SE of MEC networks. By exploiting superposition coding at transmitter and successive interference cancellation (SIC) at receiver, NOMA allows multiple users to share the same bandwidth simultaneously in either power domain or code domain to improve spectral usage \cite{hj1}. 
Extensive works have formulated EE optimization frameworks and designed energy-efficient resource allocation schemes in NOMA enabled MEC networks \cite{ee1}-\cite{ee3}.

In order to maximize EE under the delay constraints in MEC, a joint radio and computation resource allocation problem was formulated in \cite{ee1}, which takes both intra and inter cell interference into consideration.
In \cite{ee2}, the authors proposed a scheme to maximize the system computation EE of a wireless power transfer enabled NOMA based MEC network by jointly optimizing the computing frequencies, execution time, and transmit power.
In \cite{ee3}, the task offloading and resource allocation problem in a NOMA assisted MEC network with security and EE was investigated. A dynamic task offloading and resource allocation algorithm was proposed based on Lyapunov optimization theory.

However, complex wireless environment and devices' limited transmit power can render it very challenging for the NOMA based MEC network to achieve high offloading rate and high EE. IRS recently has emerged as a promising technique to tackle wireless capacity and energy issues. IRS consists of a large number of  low-cost passive reflecting elements with the adjustable phase shifts \cite{irs1}. By properly adjusting the phase shifts of the elements, the reflected signals from various paths can be combined coherently to enhance the link achievable rate at the MEC receiver \cite{irsqun}. Through this, IRS can compensate for the pathloss and fading  and allow users  to use a relatively lower power to achieve a higher data rate. Since IRS does not employ any transmit radio frequency (RF) chains, energy consumption solely comes from reflective element phase adjustment and is usually very low.  It  becomes a very promising technology to improve EE in future wireless networks.

The following studies have investigated the performance of using IRS with NOMA. 
In \cite{nomairs1}, an IRS-assisted uplink NOMA system was considered to maximize the sum rate of all the users under the individual power constraint. The considered problem requires a joint power control at the users and beamforming design at the IRS, and an SDR-based solution has been developed. 
In \cite{nomairs2}, the problem of joint user association, subchannel assignment, power allocation, phase shifts design, and decoding order determination was formulated for maximizing the sum-rate for an IRS-assisted NOMA network.
In \cite{nomairs3}, an EE algorithm was proposed to yield a tradeoff between the rate maximization and power minimization for an IRS-assisted NOMA network. The authors aimed to maximize the system EE by jointly optimizing the transmit beamforming and the reflecting beamforming. It was shown that NOMA can improve EE compared to OMA.

Furthermore, recently  application of IRS into NOMA-based MEC networks has been studied.
In \cite{nomairs4}, the authors investigated an IRS-aided MEC system with NOMA. By jointly optimizing the passive phase shifters, the size of transmission data, transmission rate, power, and time, as well as the decoding order, they aimed to minimize the sum energy consumption. A block coordinate descent method was developed to alternately optimize two separated subproblems. 
In \cite{nomairs5}, an IRS-aided MEC system was considered and a flexible time-sharing NOMA scheme was proposed to allow users to divide their data into two parts that are transmitted via NOMA and TDMA respectively. By designing the IRS passive reflection and users’ computation-offloading scheduling, the delay was minimized.

However, neither \cite{nomairs4}  nor \cite{nomairs5} considered the EE performance of IRS-assisted MEC networks with NOMA, which is very important for system design to obtain the optimal trade-off between achievable rate and consumed power.
Motivated by the above-mentioned observations, in this paper, the EE maximization problem is studied in an IRS-assisted MEC network with NOMA. To the authors' best knowledge, this is the first work that focuses on EE performance for applying both NOMA and IRS in the MEC network. The major contributions of this paper are summarized as follows.

We investigate the joint design of the receiver beamforming, offloading power, phase shift matrix, and local computing frequency to maximize the EE in an IRS-assisted MEC network with NOMA. The problem is challenging to solve due to its non-convexity fractional objective function and coupling of the beamforming vector with the IRS phase shift matrix. An alternating optimization algorithm is proposed to solve the non-convex fractional problem by using semidefinite programming relaxation (SDR).
The simulation results show that the proposed method can achieve the highest EE among all the benchmarks.

The rest of the paper is organized as follows. The system model is provided in Section II. The EE maximization problem and its solution are presented in Section III. Simulation results are given in Section IV. The paper is concluded in Section V.

\textit{Notation:} $\mathbb{C}^{M\times N}$ denotes the $M \times N $ complex-valued matrices space. $\mathcal{CN}(\mu, \sigma^2)$ denotes the distribution of complex Gaussian random variable with mean $\mu$ and variance $\sigma^2$. For a square matrix $\mathbf{X}$, the trace of $\mathbf{X}$ is denoted as $\text{Tr}(\mathbf{X})$ and rank($\mathbf{X}$) denotes the rank of matrix $\mathbf{X}$. $\angle(x)$ denotes the phase of complex number $x$. Matrices and vectors are denoted by boldface capital letters and boldface lower case letters. 

\section{System Model}
\begin{figure}[h]
	\setlength{\abovedisplayskip}{3pt}
	\setlength{\belowdisplayskip}{3pt}
	\centering
	\includegraphics[width=3.0in]{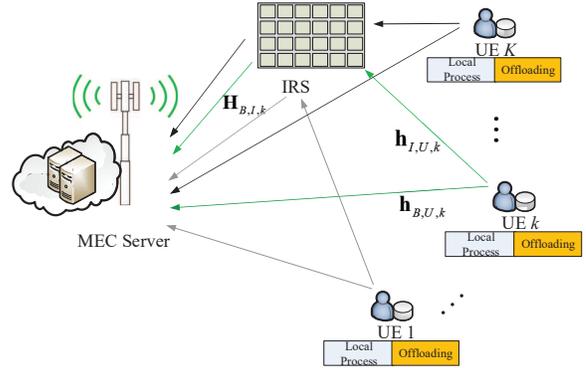}	
	\caption{An IRS-aided MEC system with NOMA.\label{model}}
\end{figure}

As shown in Fig. \ref{model}, an IRS-assisted MEC system is considered. There are $K$ single-antenna user equipments (UEs) in the system, which can do both local computing and data offloading. The access point (AP) with an MEC server is equipped with  $N$ antennas and the IRS has $M$ reflecting elements.  
\subsection{Offloading Model}
The baseband equivalent channel from UE $k$ to IRS, IRS to AP, and UE $k$ to AP are denoted as  ${\bf{h}}_{I,U,k} \in \mathbb{C}^{1 \times M}$, ${\bf{H}}_{B,I,k} \in \mathbb{C}^{M \times N}$, and ${\bf{h}}_{B,U,k} \in \mathbb{C}^{1 \times N}$, respectively. 
In this paper, IRS adjusts its elements to maximize the combined incident signal from each UE to the AP. The diagonal phase-shift matrix can be denoted as $\mathbf{\Theta} = \text{diag}(\exp({j\theta_{1}}), \exp({j\theta_{2}}), $ $\cdots, \exp({j\theta_{M}}))$, wherein its main diagonal, $\theta_{m}\in [0,2\pi)$, denotes the phase shift on the combined incident signal by its $m$th element, $m=1,2,...,M$ \cite{cwh3}.

The transmitted signal from UE $k$ is given as $\sqrt{p_k} s_k$, where $\sqrt{p_k}$ denotes the transmit power and $s_k$  denotes the independent information. $\mathbf{m}_{B,k}\in \mathbb{C}^{N \times 1}$ denotes the receive beam vectors with unit norm, i.e., $\left\| \mathbf{m}_{B,k} \right\|^2=1$ \cite{nomairs1}.
Therefore, the signal received at AP can be given as 
\be
	\setlength{\abovedisplayskip}{3pt}
\setlength{\belowdisplayskip}{3pt}
\mathbf{y}_{B,U}=\sum_{k=1}^K({\bf{h}}_{B,U,k}^H + {\bf{h}}_{I,U,k}^H\Theta {{\bf{H}}_{B,I,k}}){{\bf{m}}_{B,k}}\sqrt {{p_k}} {s_k} + n_{B,U,k},
\ee
where $n_{B,U,k} \sim \mathcal{CN}(0,\sigma^2)$ is the complex additive white Gaussian noise (AWGN) \cite{nomairs3}, \cite{chanel1}. 

NOMA is used to improve SE and mitigate the interference between different UEs. By exploiting the SIC techniques, the received signal at AP is sequentially decoded and the UE with the best channel conditions is firstly decoded. The  channel of each UE includes a direct link and a reflect link. Since the reflect link depends on the unknown parameters $\bf{\Theta}$, the effective channels cannot be used to order the users at the receiver side. Similar to \cite{nomairs1}, we simply remove unknown reflect matrix by considering  it as an identity matrix $\mathbf{I}$. UEs are then sorted based on this channel gain $|(\mathbf{h}_{B,U,k}^{H}+\mathbf{h}_{I,U,k}^{H}\mathbf{I}\mathbf{H}_{B,I,k})|$. Without loss of generality, we assume that UEs are sorted in an increasing order, i.e., $|(\mathbf{h}_{B,U,1}^{H}+\mathbf{h}_{I,U,1}^{H}\mathbf{I}\mathbf{H}_{B,I,1})|\le |(\mathbf{h}_{B,U,2}^{H}+\mathbf{h}_{I,U,2}^{H}\mathbf{I}\mathbf{H}_{B,I,2})|\le \cdots \le|(\mathbf{h}_{B,U,K}^{H}+\mathbf{h}_{I,U,K}^{H}\mathbf{I}\mathbf{H}_{B,I,K})|$. When decoding the signal for UE $k$, the signals from $i=1,2,\cdots, k-1$ are treated as interference.
Thus, the signal to interference plus noise ratio (SINR) for UE $k$ is expressed as 
\be
	\setlength{\abovedisplayskip}{3pt}
\setlength{\belowdisplayskip}{3pt}
\gamma_{B,k} = \frac{p_k|(\mathbf{h}_{B,U,k}^{H}+\mathbf{h}_{I,U,k}^{H}\mathbf{\Theta}\mathbf{H}_{B,I,k})\mathbf{m}_{B,k}|^2}{\sum_{i=1}^{k-1}p_i|(\mathbf{h}_{B,U,i}^{H}+\mathbf{h}_{I,U,i}^{H}\mathbf{\Theta}\mathbf{H}_{B,I,i})\mathbf{m}_{B,i}|^2+\sigma^2}.
\ee
The achievable offloading rate is 
\be
	\setlength{\abovedisplayskip}{3pt}
\setlength{\belowdisplayskip}{3pt}
R^{off}_{k}=B\log_2(1+\gamma_{B,k}).
\ee

\subsection{Local Processing Model}
Let $C_{k} $ be the number of computation cycles required to process one bit of data for UE $k$ locally. UE can compute and transmit simultaneously. Let $f_{k}$ denote the computing frequency of the processor (cycles/second) \cite{hj1}. Therefore, the local computing rate can be given as
\be
	\setlength{\abovedisplayskip}{3pt}
\setlength{\belowdisplayskip}{3pt}
R_{k}^{loc}=\frac{f_{k}}{C_{k}}.
\ee
The power consumption of local computing is modeled as a function of processor speed $f_{k}$. It can  be given as $p_{k}^{loc}= \epsilon f_{k}^3$, where $\epsilon$ is effective capacitance coefficient of processor chip.
\subsection{Energy Efficiency}
The energy consumed by each UE consists of transmit power, local computing power, and  circuit power consumption. 
Thus, the total power consumed by each UE is given as
\be
	\setlength{\abovedisplayskip}{3pt}
\setlength{\belowdisplayskip}{3pt}
P^{tot}_k=  p_k+\epsilon f_{k}^3+P^{cn}_k,
\ee
where $P^{cn}_k$ denotes the constant circuit power consumed for signal processing and it is assumed to be the same for all UEs.
The total achievable rate for each UE is
\be
	\setlength{\abovedisplayskip}{3pt}
\setlength{\belowdisplayskip}{3pt}
R_k^{tot}=R_k^{off}+R_k^{loc}.
\ee
According to \cite{cwh3},  EE is defined as
\be
	\setlength{\abovedisplayskip}{3pt}
\setlength{\belowdisplayskip}{3pt}
\eta =\frac{\sum_{k=1}^{K}R_k^{tot}}{\sum_{k=1}^{K}P_k^{tot}}.
\ee
In order to maximize the EE, the local CPU frequency, offloading power, decoding vectors, and the phase shift matrix need to be jointly optimized. 

\section{Resource Optimization}
In this section, the EE maximization problem is studied by jointly optimizing the local CPU frequency, offloading power, decoding vectors, and phase shift matrix. An alternating algorithm is further proposed to tackle the formulated  problem.
\vspace{-0.03in}
\subsection{Problem Formulation}
The EE maximization problem is formulated as 
\begin{subequations}
		\setlength{\abovedisplayskip}{3pt}
	\setlength{\belowdisplayskip}{3pt}
	\label{P0}
	\begin{alignat}{5}
	\textbf{P}{_1:}~ &\max_{p_k,f_k,\mathbf{m}_{B,k},\mathbf{ \Theta}}~\eta\nonumber\\
	s.t.~~&P_k^{tot} \le P^{th}_k,\\
	&R_k^{tot} \ge R_{th},\\
	&|\exp({j\theta_{m}})|=1,\\
	&\left\| \mathbf{m}_{B,k} \right\|^2=1,
	\end{alignat}
\end{subequations} 
where $R_{th}$ is the minimum required rate threshold. $P_k^{th}$ is the maximum available power of each UE. It is evident that problem $\textbf{P}_1$ is non-convex due to the fractional structure of the objective function and the non-convex constraints. In order to tackle it, an alternating algorithm is proposed.

By introducing ${\mathbf{w}^H}=[w_1, w_2, \cdots, w_M]$,  one has $\mathbf{h}_{I,U,k}^{H}\mathbf{\Theta}\mathbf{H}_{B,I,k}={\mathbf{w}^H}\mathbf{H}{_{B,k}}$, where $w_m=\exp({j\theta_{m}})$, $\mathbf{H}_{B,k}=\text{diag}(\mathbf{h}_{I,U,k}^H)\mathbf{H}_{B,I,k}$.   
Thus, the SINR of UE $k$ is given as $\gamma_{B,k}=\frac{a_0p_k|\overline{\mathbf{w}}^H\overline{\mathbf{H}}_{B,k}\mathbf{m}_{B,k}|^2}{a_0\sum_{i=1}^{k-1}p_i|\overline{\mathbf{w}}^H\overline{\mathbf{H}}_{B,i}\mathbf{m}_{B,i}|^2+1}$, where $a_0=1/\sigma^2$, $\overline{\mathbf{H}}_{B,k}=\left[ {\begin{array}{*{20}{c}}
	{{\mathbf{H}_{B,k}}}  \\
	{{\mathbf{h}_{B,U,k}}}  \\
	\end{array}} \right]$,  $\overline{\mathbf{w}}^H=\exp({j\overline{w}})[\mathbf{w}^H,1]$,  and $\overline{w}$ is an arbitrary phase rotation.
The objective of the optimization problem can be transformed into 
\be
	\setlength{\abovedisplayskip}{3pt}
\setlength{\belowdisplayskip}{3pt}
	\frac{\frac{B}{\ln2}(\ln(1+\sum_{k=1}^K{a_0p_k|\overline{\mathbf{w}}^H\overline{\mathbf{H}}_{B,k}\mathbf{m}_{B,k}|^2}))+\sum_{k=1}^KR_k^{loc}}{\sum_{k=1}^KP_k^{tot}}.
	\label{10}
\ee
 To tackle the complexity introduced by the logarithmic function of $R_k^{off}$ in (\ref{P0}\text{b}),  Lemma 1 is introduced. First, we have
\be
	\setlength{\abovedisplayskip}{3pt}
\setlength{\belowdisplayskip}{3pt}
	\begin{split}
R_k^{off}&=\frac{B}{\ln2}[\ln(a_0\sum_{i=1}^{k}p_i|\overline{\mathbf{w}}^H\overline{\mathbf{H}}_{B,i}\mathbf{m}_{B,i}|^2+1)\\
&-\ln({a_0\sum_{i=1}^{k-1}p_i|\overline{\mathbf{w}}^H\overline{\mathbf{H}}_{B,i}\mathbf{m}_{B,i}|^2+1})].
\end{split}
\ee
$\mathbf{Lemma~1}$: By introducing the function $\phi(t)=-tx+\ln t+1$ for any $x > 0$, one has
\be
	\setlength{\abovedisplayskip}{3pt}
\setlength{\belowdisplayskip}{3pt}
-\ln x=\max_{t>0}\phi(t).
\ee
The optimal solution can be achieved at $t=1/x$. 
By setting $x={a_0}\sum\limits_{i = 1}^{k - 1} {{p_i}} |{\overline {\bf{w}} ^H}{\overline {\bf{H}} _{B,i}}{{\bf{m}}_{B,i}}{|^2} + 1$, and $t=t_{B,k}$, one has 
\be
	\setlength{\abovedisplayskip}{3pt}
\setlength{\belowdisplayskip}{3pt}
\begin{split}
	R_k^{off}&=\frac{B}{\ln2}\max_{t_{B,k}>0}\phi_{B,k}(p_k,f_k,\mathbf{m}_{B,k},\overline{\mathbf{w}},t_{B,k})\\
	&=\frac{B}{\ln2}[\ln(a_0\sum_{i=1}^{k}p_i|\overline{\mathbf{w}}^H\overline{\mathbf{H}}_{B,i}\mathbf{m}_{B,i}|^2+1)+ \ln ({t_{B,k}}) + 1\\
	& - {t_{B,k}}({a_0}\sum\limits_{i = 1}^{k - 1} {{p_i}} |{\overline {\bf{w}} ^H}{\overline {\bf{H}}_{B,i}}{{\bf{m}}_{B,i}}{|^2} + 1) ].	
\end{split}
\ee
By further introducing a variable $\eta_1$ to deal the fractional structure of (\ref{10}), $\textbf{P}{_{1}}$ can be transformed into
\begin{subequations}
		\setlength{\abovedisplayskip}{3pt}
	\setlength{\belowdisplayskip}{3pt}
	\begin{alignat}{5}
		\label{p3}
		\begin{split}
			\textbf{P}_{2}:~ &\max_{p_k,f_k,\mathbf{m}_{B,k},t_{B,k},\overline{\mathbf{w}}}~ [\sum_{k=1}^KR_k^{loc}+\frac{B}{\ln2}(\ln(1+\\
			&\sum_{k=1}^K{a_0p_k|\overline{\mathbf{w}}^H\overline{\mathbf{H}}_{B,k}\mathbf{m}_{B,k}|^2}))]-\eta_{1}{\sum_{k=1}^KP_k^{tot}}
		\end{split}\nonumber\\
		s.t.~~& (\ref{P0}\text{a}), (\ref{P0}\text{c}),(\ref{P0}\text{d}),\nonumber\\\
		\begin{split}
			&\frac{B}{\ln2}\phi_{B,k}(p_k,f_k,\mathbf{m}_{B,k},\overline{\mathbf{w}},t_{B,k})+R_k^{loc} \ge R_{th}.
		\end{split}
	\end{alignat}
\end{subequations}

$\textbf{P}_{2}$ is still non-convex due to the coupling of variables.
An alternating algorithm is proposed. To be specific, $p_k$ and $f_k$ are first optimized with a given $\mathbf{m}_{B,k}$, and $\overline{\mathbf{w}}$. $\mathbf{m}_{B,k}$ can then be optimized with the obtained $p_k,f_k$,  and $\overline{\mathbf{w}}$. Further $\overline{\mathbf{w}}$ can be optimized with the obtained $p_k,f_k$, and $\mathbf{m}_{B,k}$. This process iteratively continues until convergence.  
\subsection{CPU Frequency and Offloading Power Optimization}
With the given $\mathbf{m}_{B,k}$ and $\overline{\mathbf{w}}$, let $A_{B,k}=a_0|\overline{\mathbf{w}}^H\overline{\mathbf{H}}_{B,k}\mathbf{m}_{B,k}|^2$, the problem can be transformed into
\begin{subequations}
		\setlength{\abovedisplayskip}{3pt}
	\setlength{\belowdisplayskip}{3pt}
	\begin{alignat}{5}
		\begin{split}
			\textbf{P}{_{3.1}:}~ &\max_{p_k,f_k}~\frac{B}{\ln2}(\ln(\sum_{k=1}^{K}p_kA_{B,k}+1))\\
			&+\sum_{k=1}^{K}\frac{f_k}{C_k}-\eta_1 \sum_{k=1}^K(\zeta p_k+\epsilon f_{k}^3+P^{cn}_k)
		\end{split}\nonumber\\
		s.t.~~&  p_k+\epsilon f_{k}^3+P^{cn}_k \le P_k^{th},\\
		\begin{split}
			&\frac{B}{\ln2}\phi_{B,k}(p_k,f_k)+\frac{f_k}{C_k} \ge R_{th}.
		\end{split}
	\end{alignat}
\end{subequations}
Problem $\textbf{P}_{3.1}$ is convex with respect to $f_k$ and $p_k$, therefore, it can be solved by using a standard convex optimization tool. 

\subsection{Optimizing the Receiving Beamforming}
	\setlength{\abovedisplayskip}{3pt}
\setlength{\belowdisplayskip}{3pt}
In this section, we solve the problem $\textbf{P}_{2}$ to achieve the receive beamforming vector $\mathbf{m}_{B,k}$ for a given $\overline{\mathbf{w}}$, $p_k$, and $f_k$.
Let $\mathbf{\overline{h}}_{B,k}^H=\mathbf{\overline{w}}^H\mathbf{\overline{H}}_{B,k}$, $\textbf{P}_{2}$ can be transformed into 
\begin{subequations}
		\setlength{\abovedisplayskip}{3pt}
	\setlength{\belowdisplayskip}{3pt}
	\begin{alignat}{5}
		\label{p3.2}
		\begin{split}
			\textbf{P}_{3.2}:~ &\max_{\mathbf{m}_{B,k}}~\frac{\ln2}{B}\ln(a_0\sum_{k=1}^{K}p_k|\mathbf{\overline{h}}_{B,k}^H\mathbf{m}_{B,k}|^2\\
			&+1)+\sum_{k=1}^KR_k^{loc}-\eta_{1}\sum_{k=1}^KP_k^{tot}
		\end{split}\nonumber\\
		\begin{split}
			s.t.~~&\frac{B}{\ln2}\ln(a_0\sum_{i=1}^{k}p_i|\mathbf{\overline{h}}_{B,k}^H\mathbf{m}_{B,i}|^2+1)+ \ln ({t_{B,k}}) + 1\\
			& - {t_{B,k}}({a_0}\sum\limits_{i = 1}^{k - 1} {{p_i}} |{\mathbf{\overline{h}}_{B,i}^H{{\bf{m}}_{B,i}}}{|^2} + 1) +R_k^{loc}\ge R_{th},
		\end{split}\\
	&\left\| \mathbf{m}_{B,k} \right\|^2=1.
	\end{alignat}
\end{subequations}
Let $|\overline{\mathbf{h}}_{B,k}^H\mathbf{m}_{B,k}|^2 = \rm{Tr}(\tilde{\mathbf{H}}_{B,k}\mathbf{m}_{B,k}\mathbf{m}_{B,k}^H)$. By defining $\tilde{\mathbf{H}}_{B,k}=\mathbf{\overline{h}}_{B,k}\mathbf{\overline{h}}_{B,k}^H$,  $\mathbf{M}_{B,k} = \mathbf{m}_{B,k}\mathbf{m}_{B,k}^H$, one has $\mathbf{M}_{B,k}\succeq0$ and $\text{rank}(\mathbf{{M}}_{B,k})=1$. 
The rank-$1$ constraint makes  the problem difficult to solve. Thus, we apply the SDR method to relax the constraints \cite{irsqun}. $\textbf{P}_{3.2}$ is then expressed as 
\begin{subequations}
		\setlength{\abovedisplayskip}{3pt}
	\setlength{\belowdisplayskip}{3pt}
	\begin{alignat}{5}
		\label{p3.3}
		\begin{split}
			\textbf{P}_{3.3}:~ &\max_{\mathbf{M}_{B,k}}~[\frac{B}{\ln2}\ln(a_0\sum_{k=1}^{K}p_k\rm{Tr}(\tilde{\mathbf{H}}_{B,k}{\mathbf{M}_{B,k}})\\
			&+1) +\sum_{k=1}^KR_k^{loc}]-\eta_{1}\sum_{k=1}^KP_k^{tot}
		\end{split}\nonumber\\
		\begin{split}
			s.t.~~&\frac{B}{\ln2}\ln(a_0\sum_{i=1}^{k}p_i\rm{Tr}(\tilde{\mathbf{H}}_{B,i}{\mathbf{M}_{B,i}})+1)+ \ln ({t_{B,k}}) + 1\\
			& - {t_{B,k}}({a_0}\sum\limits_{i = 1}^{k - 1} {{p_i}} \rm{Tr}(\tilde{\mathbf{H}}_{B,i}{\mathbf{M}_{B,i}}) + 1) +R_k^{loc}\ge R_{th},
		\end{split}\\
	&\rm{Tr}({\mathbf{M}_{B,k}})=1.
	\end{alignat}
\end{subequations}
$\textbf{P}_{3.3}$ is convex and can be solved by using a standard convex optimization tool \cite{wqqirs}. 
After $\mathbf{M}_{B,k}$ is obtained, if $\text{rank}(\mathbf{M}_{B,k})=1$,  $\mathbf{m}_{B,k}$ can be obtained from $\mathbf{M}_{B,k}=\mathbf{m}_{B,k}\mathbf{m}_{B,k}^H$ by performing the eigenvalue decomposition. Otherwise, the Gaussian randomization can be used for recovering $\mathbf{m}_{B,k}$ \cite{wqqirs}. 

\subsection{Optimizing the IRS Reflecting Shifts $\overline{\mathbf{w}}$}
After obtaining the beamforming vectors $\mathbf{m}_{B,k}$, by setting $\mathbf{h}_{W,B,k}=\mathbf{\overline{H}}_{B,k}\mathbf{m}_{B,k}$, problem $\textbf{P}_{2}$ can be transformed into 
\begin{subequations}
		\setlength{\abovedisplayskip}{3pt}
	\setlength{\belowdisplayskip}{3pt}
	\begin{alignat}{5}
		\begin{split}
			\textbf{P}{_{3.4}:}~ &\max_{\overline{\mathbf{w}}}~\frac{B}{\ln2}\ln(a_0\sum_{k=1}^{K}p_k(\overline{\mathbf{w}}^H\mathbf{h}_{W,B,k})+1)\\
			& +\sum_{k=1}^K\frac{f_k}{C_k} -\eta_{1}\sum_{k=1}^KP_k^{tot}
		\end{split}\nonumber\\
		s.t.~~& |w_{m}|=1,~m=1,2,...,M,\\
		\begin{split}
			&\ln(a_0\sum_{i=1}^{k}p_i(\overline{\mathbf{w}}^H\mathbf{h}_{W,B,i})+1)+ \ln ({t_{B,k}}) + 1 \\
			&- {t_{B,k}}({a_0}\sum\limits_{i = 1}^{k - 1} {{p_i}} (\overline{\mathbf{w}}^H\mathbf{h}_{W,B,i}) + 1) +\frac{f_k}{C_k} \ge R_{th}.
		\end{split}
	\end{alignat}
\end{subequations}
Similar to the previous section, let $\mathbf{W}=\overline{\mathbf{w}}\overline{\mathbf{w}}^H$, $\mathbf{H}_{W,B,k}=\mathbf{h}_{W,B,k}\mathbf{h}_{W,B,k}^H$. By applying the SDR method, we have 
\begin{subequations}
		\setlength{\abovedisplayskip}{3pt}
	\setlength{\belowdisplayskip}{3pt}
	\begin{alignat}{5}
		\begin{split}
			\textbf{P}{_{3.5}:}~ &\max_{\mathbf{W}}~\frac{B}{\ln2}\ln(a_0\sum_{k=1}^{K}p_k\text{Tr}(\mathbf{H}_{W,B,k}\mathbf{W})+1)\\
			& +\sum_{k=1}^K\frac{f_k}{C_k} -\eta_{1}\sum_{k=1}^KP_k^{tot}
		\end{split}\nonumber\\
		s.t.~~& \mathbf{W} \succeq 0, \mathbf{W}_{mm}=1,~m=1,2,...,M,\\
		\begin{split}
			&\ln(a_0\sum_{i=1}^{k}p_i\text{Tr}(\mathbf{H}_{W,B,k}\mathbf{W})+1)+ \ln ({t_{B,k}}) + 1 \\
			&- {t_{B,k}}({a_0}\sum\limits_{i = 1}^{k - 1} {{p_i}} \text{Tr}(\mathbf{H}_{W,B,k}\mathbf{W}) + 1)+\frac{f_k}{C_k} \ge R_{th}.
		\end{split}
	\end{alignat}
\end{subequations}
The problem $\textbf{P}_{3.5}$ is a convex problem and can be solved by using a standard convex optimization tool.
After obtaining $\mathbf{W}$, $\overline{\mathbf{w}}$ can be given by eigenvalue decomposition if $\text{rank}({\mathbf{W}})=1$; otherwise, the Gaussian randomization can be used for recovering the approximate $\overline{\mathbf{w}}$ \cite{wqqirs}. The reflection coefficients can be given by $w_m=\angle(\frac{\overline{{w}}_m}{\overline{{w}}_{M+1}}),~m=1,2,..,M$.
The overall optimization algorithm is summarized in Algorithm 1, where $\delta$ is the threshold and $T$ is the maximum number of iterations.
 \begin{table}[htbp]
 	\setlength{\abovedisplayskip}{3pt}
 	\setlength{\belowdisplayskip}{3pt}
 	\centering
 	\begin{center}
 		\begin{tabular}{lcl}
 			\\\toprule
 			{	$\textbf{Algorithm 1}$: Alternating Algorithm for Solving $\textbf{P}_{1}$}\\ \midrule
 			\  1) \textbf{Input settings:}\\
 			\ \ \ \ \ \ \ $\delta$, $R_{th},P_k^{th} >0$, and $T$.\\
 			\  2) \textbf{Initialization:}\\
 			\ \ \ \ \ \ \ $t_{B,k}(0)$, $\mathbf{\overline{w}}(0)$,$\mathbf{m}_{B,k}(0)$, and $\eta_1(0)$;\\
 			\  3) \textbf{Optimization:}\\
 			\ \ \ \ \ \textbf{$\pmb{\unrhd} $  for \emph{$\tau_1$}=1:T }\\
 			\ \ \ \ \ \ \ \ \ \ solve $\textbf{P}_{3.1}$ with $\mathbf{\overline{w}}^{*}(\tau_1-1)$,$\mathbf{m}_{B,k}^*(\tau_1-1)$;\\
 			\ \ \ \ \ \ \ \ \ \ obtain the solution $p_k^*(\tau_1)$, $f_k^*(\tau_1)$;\\
 			\ \ \ \ \ \ \ \ \ \ solve $\textbf{P}_{3.3}$ with $p_k^*(\tau_1)$, $f_k^*(\tau_1)$, and $\mathbf{\overline{w}}^{*}(\tau_1-1)$;\\
 			\ \ \ \ \ \ \ \ \ \ obtain the solution $\mathbf{m}_{B,k}^*(\tau_1)$;\\
 			\ \ \ \ \ \ \ \ \ \ solve $\textbf{P}_{3.5}$ with $p_k^*(\tau_1)$, $f_k^*(\tau_1)$, and $\mathbf{m}_{B,k}^*(\tau_1)$;\\
 			\ \ \ \ \ \ \ \ \ \ obtain the solution $\mathbf{\overline{w}}^{*}(\tau_1)$;\\
 			\ \ \ \ \ \ \ \ \ \ calculate EE $\eta(\tau_1)$ and update $t_{B,k}(\tau_1)$ and $\eta_1(\tau_1)$;\\			
 			\ \ \ \ \ \ \ \ \ \  \textbf{if} $| \frac{\eta(\tau_1)-\eta(\tau_1-1)}{\eta(\tau_1)}| \le \delta$; \\
 			\ \ \ \ \ \ \ \ \ \ \ \ \ \ the optimal EE $\eta^*$ is obtained;\\
 			\ \ \ \ \ \ \ \ \ \  \textbf{end}\\
 			\ \ \ \ \ \textbf{$\pmb{\unrhd} $ end} \\
 			\  4) \textbf{Output:}\\
 			\ \ \ \ \ \ \ $p_k^*$, $f_k^*$, $\mathbf{m}_{B,k}^*$, and $\mathbf{w}^{*}$ and EE $\eta^*$.\\
 			\bottomrule
 		\end{tabular}
 	\end{center}
 \end{table}
\section{Simulation Results}
In this section, simulation results are provided to evaluate the performance of the proposed algorithms. We consider a three-dimensional Cartesian coordinate system. The simulation settings are based on those used in \cite{irs1}, \cite{wqqirs}. 
We consider a $2$-UE case and it can be readily extend to multiple UE cases. The locations of the MEC, the IRS, UE1, and UE2 are set as $(5,0,20)$, $(0,50,2)$, $(5,75,5)$ and $(5,50,10)$, respectively \cite{wqqirs}.
The channels are generated by  $h_{i,j}=\sqrt{G_0d_{i,j}^{-c_{i,j}}}g_{i,j}$, where $G_0=-30$ dB is the path loss at the reference point. $d_{i,j}$, $c_{i,j}$ and $g_{i,j}$ denote the distance, path loss exponent, and fading between $i$ and $j$, respectively, where $i \in \{B, I\}$ and $j \in \{U,k\}$. The path loss exponents are set as $c_{B,U,k}=5$, $c_{B,I}=3.5$, and $c_{I,U,k}=2$. 
The bandwidth $B$ is set to $1$ Mhz. Other parameters are set as $\sigma^2= -105$ dBm, $P_k^{th}=31$ dBm, $P_k^{cn}=23$ dBm, $C_k=10^3$ cycles/bit, and $\epsilon=10^{-28}$.

The  proposed scheme is marked as `Efficiency-IRS'. We consider three other cases as benchmarks to compare with the proposed method. The first benchmark, marked as `OMA-IRS', uses FDMA with equally allocated bandwidth to all the users. The second benchmark, marked as `OnlyOff-IRS',  has no local computing and all the tasks are offloaded. The third benchmark, marked as `Efficiency-NoIRS", aims to investigate the performance without IRS.

\begin{figure}[h]
		\setlength{\abovecaptionskip}{-0.2cm} 
	\setlength{\belowcaptionskip}{-1cm}
	\centering
	\includegraphics[width=3.0in]{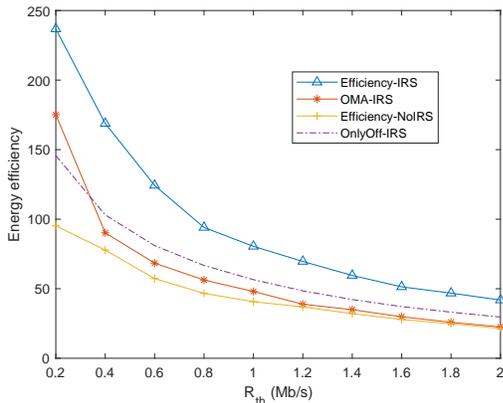}	
\caption{Energy efficiency versus the minimum rate threshold.\label{etarth}}

\end{figure}	
Fig. \ref{etarth} shows EE versus the minimum rate threshold $R_{th}$. 
The EE achieved by the proposed method is the best among all the schemes.  This indicates that the IRS assisted MEC with NOMA can help improve the system rate and achieve high EE. 
With the increase of $R_{th}$, all the curves are decreasing. The system has to consume excessive energy to increase the rate in order to meet the minimum rate constraint, which decreases  EE.

\begin{figure}[h]
		\setlength{\abovecaptionskip}{-0.2cm} 
	\setlength{\belowcaptionskip}{-1cm}
	\centering
	\includegraphics[width=3.0in]{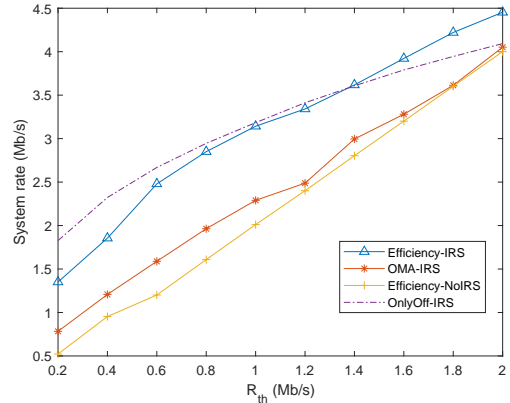}	
	\caption{Achievable rate versus the minimum rate threshold.\label{raterth}}
\end{figure}

A comparison of the system rate versus the rate threshold $R_{th}$ is presented in Fig. \ref{raterth}. All the curves increase with $R_{th}$ in order to meet the service requirement, which verifies the observation in Fig. \ref{etarth}. 
The system rates obtained by the proposed method and the `OnlyOff-IRS' method are higher than those of the other two methods, which indicates that combining IRS with NOMA can significantly help the system to achieve a higher rate. It is worth noting that even though the `OnlyOff-IRS' method can achieve the highest rate when $R_{th}$ is low, its efficiency is lower than the proposed method. This indicates that the overall efficiency performance degrades when there is no local computing.
\begin{figure}[h]
		\setlength{\abovecaptionskip}{-0.2cm} 
	\setlength{\belowcaptionskip}{-1cm}
	\centering
	\includegraphics[width=3.0in]{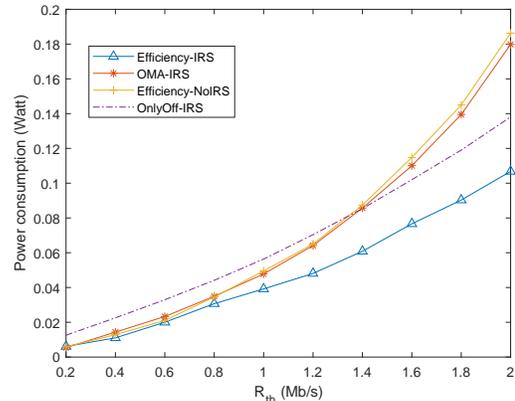}	
	\caption{Power consumption versus the minimum rate threshold.\label{powerrth}}
\end{figure}
\begin{figure}[h]
			\setlength{\abovecaptionskip}{-0.2cm} 
	\setlength{\belowcaptionskip}{-1cm}
	\centering
	\includegraphics[width=3.0in]{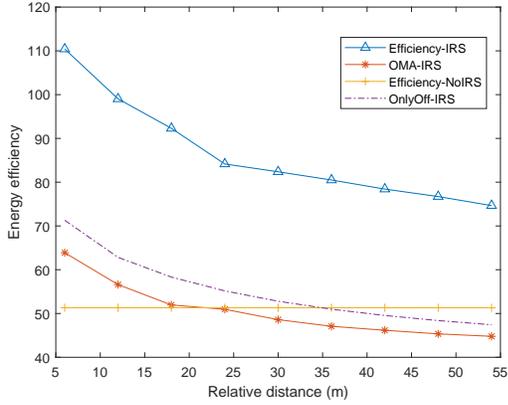}	
	\caption{Energy efficiency versus the relative distance of UE-IRS.\label{etairs}}
\end{figure}
\begin{figure}[h]
			\setlength{\abovecaptionskip}{-0.2cm} 
	\setlength{\belowcaptionskip}{-1cm}
	\centering
	\includegraphics[width=3.0in]{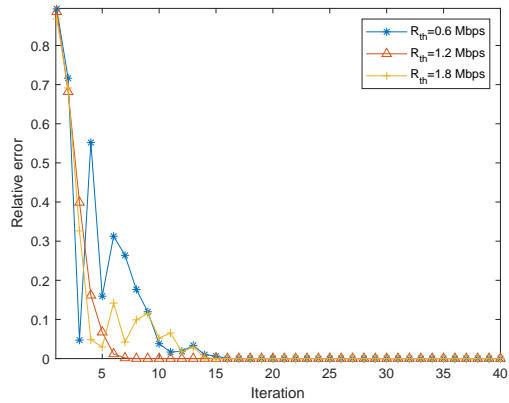}	
	\caption{Convergence with Iteration.\label{iters}}
\end{figure}

Fig. \ref{powerrth} presents the power consumption versus $R_{th}$ for different methods. The results of all the methods in Fig. \ref{powerrth} are consistent with what are shown in Fig. \ref{etarth} and Fig. \ref{raterth}. It is worth noting that the power consumption by the proposed method is quite low. So UEs can use less energy to achieve a higher rate, which demonstrates the advantage of combining NOMA and IRS to MEC network in improving EE.

Fig. \ref{etairs} shows EE versus the distance between UEs and IRS. The distance is the relative increased amount compared with  UEs' original position. The curves for all the methods with IRS decrease  with the increase of the distance except `Efficiency-NoIRS'. This is because the increase of the distance results in the increase of the path loss and the reduction of the power gain from the reflecting path through the IRS. Therefore, the achievable rate and EE are both decreased. It can also be  seen that the `Efficiency-IRS' method still has the highest performance among all the methods, which validates the superiority of the  proposed design.

In Fig. \ref{iters}, the coverage of proposed methods based on different $R_{th}$ setting are investigated. It can be observed from Fig. \ref{iters} that only several iterations are needed for the proposed algorithms to converge, showing the computation efficiency of the proposed algorithm.
\section{Conclusions}
In this paper, an IRS-assisted MEC network with NOMA was considered.  EE was maximized by jointly optimizing the offloading power, local computing frequency, beamforming vectors, and IRS phase shift matrix. An alternating algorithm was proposed to solve the challenging non-convex fractional optimization problems. The numerical results showed that our proposed method outperforms other benchmark schemes in terms of EE. It was proved that NOMA and IRS could help the MEC network to achieve a higher rate with a lower power. The convergence of the proposed algorithm was also verified. 

\end{document}